\documentclass[pra,showpacs,onecolumn,reprint,aps,]{revtex4-2}
\usepackage[utf8]{inputenc}

\usepackage{amsfonts}
\usepackage{amsmath}
\usepackage{amssymb,amscd}
\usepackage{psfrag}
\usepackage{graphicx}
\usepackage{braket}
\usepackage[dvips]{epsfig}
\usepackage{epsfig}
\usepackage{subfigure}
\usepackage{color} 
\usepackage{amssymb}
\usepackage{lineno}
\usepackage{quantikz}

\begin{document}
\nolinenumbers
\title{Temporal SU(1,1) Interferometer with Broadband Squeezed Light Injection }
\author{Zepeng Liu$^{1 }$}
\author{Tianyu Liu$^{1 }$}
\author{Hongmei Ma$^{2 }$}
%\author{Liqing Chen$^{1 }$}
%\email{lqchen@phy.ecnu.edu.cn}
%\author{Weiping Zhang$^{3,4,5,6}$}
\author{Chun-Hua Yuan$^{1}$}
\email{chyuan@phy.ecnu.edu.cn}
\date{\today }

\affiliation{$^1$State Key Laboratory of Precision Spectroscopy, Quantum
	Institute for Light and Atoms, School of Physics and Electronic Science,
	East China Normal University, Shanghai 200062,China}
\affiliation{$^2$School of Communication and Electronic Engineering, East China Normal University, Shanghai 200062, China}
%\address{$^3$School of Physics and Astronomy, and
%	Tsung-Dao Lee Institute, Shanghai Jiao Tong University,  Shanghai 200240, China}
%\address{$^4$shanghai Branch, Hefei National Laboratory, Shanghai 201315,
%	China}
%\address{$^5$Collaborative Innovation Center of Extreme Optics,
%	Shanxi University, Taiyuan, Shanxi 030006, China}
%\address{$^{6}$Shanghai
	%Research center for Quantum Science, Shanghai 201315, China}

\begin{abstract}
Temporal optics has attracted much attention due to its ability for lossless stretching of ultrafast temporal pulses. At the same time, spatial SU(1,1) interferometers have been widely used because of their high sensitivity to phase changes. On this basis, we studied a temporal SU(1,1) interferometer based on a temporal Fourier transform system and injected broadband squeezing light into the interferometer for research. The results show that the output spectral characteristics of the interferometer depend on the ratio of the focal group velocity dispersion (GDD) of the two temporal lenses (this ratio is defined as the scaling factor $M$) and the phase derivative of the applied phase. The scaling factor $M$ significantly affects the bandwidth and squeezing degree of the output spectrum. The phase derivative induces a frequency-shift effect, and the magnitude of the shift exhibits a linear relationship to the phase derivative. Furthermore, in the output squeezed-state spectrum, the distribution of squeezing degree concentrates at the center frequency and at positions where frequency shifts occur. As the value of scaling factor $M$ increases, the proportion of squeezing degree allocated at the center frequency correspondingly increases. This temporal SU(1,1) interferometer architecture opens new avenues for the control of non-classical fields in the time-frequency domain and quantum information processing applications.
\end{abstract}

\maketitle
% Your name



%\tableofcontents

\section{Introduction}
The spectral-temporal degrees of freedom inherent in optical fields have established a robust and highly promising platform for the domain of quantum information coding \cite{mk1,MK2}. In the process of quantum network construction and application, due to the huge differences and diversity of the technical means adopted \cite{MK12}, the optimization of the mode matching between different nodes is required. To achieve this goal, researchers have made continuous efforts and proposed a variety of effective strategies. For example, carrier frequency conversion technology can cleverly change the carrier frequency of the optical signal to make it compatible with other nodes \cite{MK13,MK14,Mk15,MK16}. Bandwidth compression or stretching operations at the same carrier frequency can flexibly adjust the bandwidth characteristics of the optical signal according to actual requirements, to better meet the communication demands between different nodes in the quantum network \cite{MK17,MK18,MK19}. 

In addition to the above, there is another method for manipulating light pulses that is very innovative and inspiring. It is based on the time-space analogy between diffracted beams and short pulses propagating in dispersive media \cite{mk31}. Based on the above duality, many mature optical theories and applications have corresponding contents in the time domain. The temporal imaging system is the first emerging field. Quantum time imaging (QTI) \cite{PhysRevA.98.053815}, as an extension and application of time imaging technology in the quantum field, aims to precisely manipulate the time-frequency degrees of freedom of quantum states without destroying the quantum state. The key component of QTI is the time lens, which introduces quadratic time-phase modulation into the input waveform, similar to the quadratic phase modulation introduced by the spatial lens on the input wavefront. Time lenses based on various principles such as electro-optic phase modulation \cite{MK17,MK19,mk32}, sum frequency generation (SFG) \cite{mk37,mk38} or four-wave mixing (FWM) \cite{mk39,mk40,mk41,Fourier_2} have achieved remarkable research results. These time lenses can provide extremely large time magnification, providing strong technical support for the development and application of QTI technology.

%%%%%%%%%%%%%%%%%%

Interferometers are highly sensitive to phase differences and are widely used in various schemes. The classic SU(2) interferometer is based on a linear beam splitter. The intensity distribution of its output light field depends on the relative phase difference of the input light field. Its phase sensitivity is the shot noise limit $1/{\sqrt{N}}$, where $N$ is the total number of photons in the interferometer \cite{giovannetti2004quantum}. Yurke et al. proposed the SU(1,1) interferometer by replacing the beam splitter with a nonlinear optical device. Its phase sensitivity breaks through the shot noise limit and reaches $1/N$ \cite{1986PRA_Yurke}. Recently, Moti Fridman's research group, based on temporal optics research, used a temporal Fourier transform system as a beam splitter to replace the traditional nonlinear beam splitter, extended the spatial SU(1,1) interferometer to the time domain, proposed the concept of a temporal SU(1,1) interferometer, and successfully realized experimental verification \cite{2023PRLtemporalSU11}. However, the field input-output relationship of this temporal SU(1,1) interferometer lacks a full quantum description. The temporal SU(1,1) interferometer can be better used to achieve temporal frequency control of quantum states such as squeezed states due to its high temporal resolution.

This paper theoretically studies the temporal SU(1,1) interferometer based on the temporal Fourier transform system, employing a full quantum model. The theoretical results are then applied to the injection of non-classical light, with broadband squeezed light serving as a representative example. Then, we perform homodyne measurements on the output optical field and ultimately derive the mathematical expressions between the input broadband squeezed light and the output squeezing spectrum function. The output spectral characteristics of the interferometer are influenced by the ratio of the group velocity dispersion (GDD) at the focus of the two temporal lenses and the phase derivative. The derivative of the phase induces a frequency shift effect, the shift amount being linearly related to the phase derivative. In squeezed state spectra, squeezing is predominantly distributed at the central frequency and the positions where frequency shifts occur, and as the value of parameter M increases, the proportion of squeezing at the central frequency increases accordingly.

The paper is structured as follows. In Sec.~II, We theoretically derive the general expressions for the input-output relations of the temporal Fourier transform system and the temporal SU(1,1) interferometer in the full quantum case based on the four-wave-mixing time lens. In Sec.~III, We study the temporal SU(1,1) interferometer with broadband squeezed light input and analyze how the output spectral characteristics of the interferometer are influenced by the scaling factor and the phase derivative. Finally, in Sec.~IV, we briefly summarize the work.

\section{Temporal Fourier transform System and temporal SU(1,1) interferometer}
In this section, we theoretically derive the general expressions for the input-output relations of the temporal Fourier transform system and the temporal SU(1,1) interferometer in the full quantum case based on the time lens of four-wave mixing.

\subsection{Temporal Fourier Transform System}
\begin{figure}[ht!]
\begin{center}
\centering{\includegraphics[scale=0.1,angle=0]{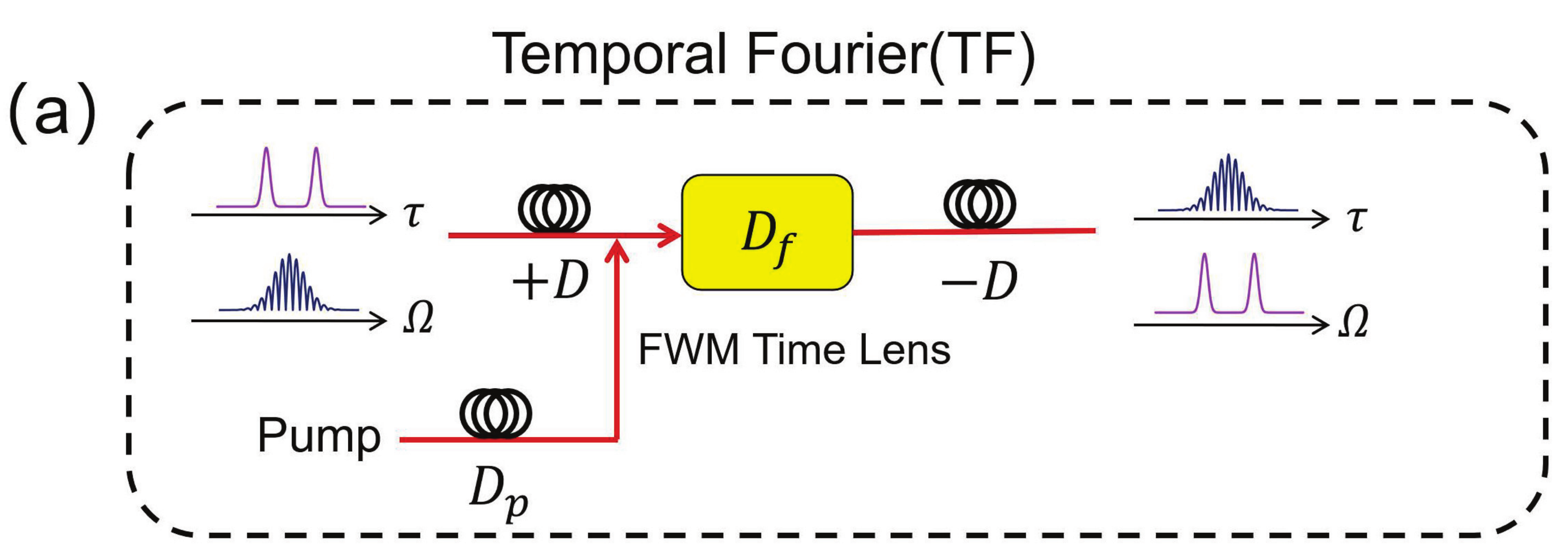}}
\centering{\includegraphics[scale=0.1,angle=0]{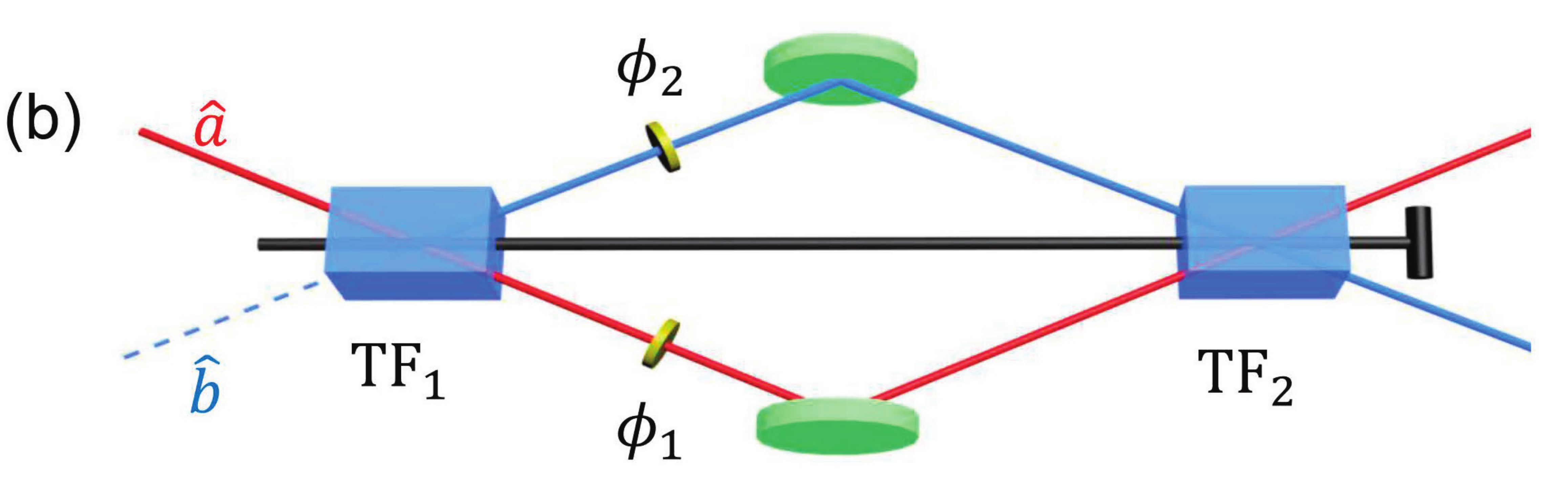}}
\end{center}
\caption{(a) Schematic diagram of temporal Fourier (TF) system. Time lens is realized by means of four-wave mixing (FWM). The chirped pump waves have a linear frequency shift as a function of time which is equivalent to a temporal quadratic phase shift.
(b) Temporal SU(1,1) interferometer. Two time Fourier transform systems (TF$_1$ and TF$_2$) are used for beam splitting and recombination. After TF$_1$, in the Fourier plane, we add $\varphi _{1} $ and $\varphi _{2}$ phases to the signal and idler.}\label{fig1}
\end{figure}

We consider a temporal Fourier transform system (TF) as shown in Fig.\ref{fig1}(a), which, like a spatial 2f system, can apply a Fourier transform \cite{Fourier_1,Fourier_2,Fourier_3} to the input signal. This is achieved by placing the input and output planes at the focal length on opposite sides of the lens, also known as a temporal $2f$ system.

The signal and idler are considered quantum mechanical operators, and the pump is considered a classical function. All waves are assumed to be narrowband, with a carrier frequency $\omega_{\mu}$, where $\mu =\left\{ s,i,p\right\} $ denotes the input signal, idler, and pump field, respectively. Each wave passing through the medium experiences dispersion, which is characterized by the dependence of the wave vector $k_{\mu}(\omega)$ on the frequency $\omega$. Expand it around $\Omega = \omega - \omega_{\mu}$ and restrict the Taylor series to the first three terms:
\begin{equation}
k_{\mu}(\omega) \approx k_{\mu}(\omega_{\mu}) + \beta_{\mu}^{(1)}\Omega + \beta_{\mu}^{(2)}\Omega^{2}/2,
\end{equation}
where $\beta_{\mu}^{(1)} = (dk_{\mu} / d\Omega)_{\omega_{\mu}}$ is the inverse of the group velocity, and $\beta_{\mu}^{(2)} = (d^{2}k_{\mu} / d\Omega^{2})_{\omega_{\mu}}$ is the group velocity dispersion of the medium at the carrier frequency $\omega_{\mu}$.

In a dispersive medium, when $\beta_{\mu}^{(2)}\neq0$, the frequency domain operator will obtain a phase shift $\hat{a}_{\mu }^{\left( 1\right) }\left( z+l,\tau \right) =\hat{a}_{\mu }^{\left( 0\right) }\left( z,\tau \right) \exp(i\beta_{\mu}^{(2)}\Omega^{2}l/2)$ during the propagation distance $l$, which can be expressed in the time domain as \cite{PhysRevA.98.053815}
\begin{equation}
    \hat{a}_{\mu }^{\left( 1\right) }\left( \tau \right) =\int_{-\infty }^{\infty
}G_{in}\left( \tau -\tau ^{\prime }\right) \hat{a}_{\mu }^{ in
}\left( \tau ^{\prime }\right) d\tau ^{\prime },
\label{shurusesan}
\end{equation}
where $G_{in}\left( \tau \right) =\frac{1}{\sqrt{-2\pi iD_{in}}}e^{-i\tau
^{2}/2D_{in}}$, $D_{in}=\beta ^{\left( 2\right) }l$ is the group delay dispersion (GDD) of the input dispersive medium.

The four-wave mixing time lenses can be realized by a nonlinear interaction of two chirped pump waves with an input signal wave. The chirped pump waves have a linear frequency shift as a function of time which is equivalent to a temporal quadratic phase shift. Under the phase matching and pump undepletion approximation, the input-output relationship of the four-wave mixing time lens is expressed as \cite{shi2017quantum}
\begin{eqnarray}
\hat{a}_{s}^{\left( 2\right) }(L, \tau ) &=&G(\tau )\hat{a}_{s}^{\left(
1\right) }(0, \tau )+ig(\tau )e^{i\phi (\tau )}\hat{a}_{i}^{\left( 1\right)
\dagger }(0, \tau), \label{eq2}
\\
\hat{a}_{i}^{\left( 2\right) }(L, \tau ) &=&ig(\tau )e^{i\phi (\tau )}\hat{a}%
_{s}^{\left( 1\right) \dagger }(0, \tau )+G(\tau )\hat{a}_{i}^{\left( 1\right)
}(0, \tau),\label{eq3}
\end{eqnarray}
where $\phi (\tau )=\phi_{1} (\tau )+\phi_{2} (\tau )=\tau ^{2}/2D_{f}$, with
parameter $D_{f}$ is the focal GDD of the time lens, which is analogous to
the focal length of the space lens. Eqs.~(\ref{eq2}) and ~(\ref{eq3})
describe the parametric amplification of signal and idler photons from the
input ($z=0$) to the output ($z=L$) of the time lens. The coefficients $%
G\left( \tau \right) $ and $g\left( \tau \right) $ is as follows 
\begin{eqnarray}
G(\tau ) &=&\cosh \left[ gA_{1}(\tau )A_{2}\left( \tau \right) L\right],
\label{eq5} \\
g(\tau ) &=&\sinh \left[ gA_{1}(\tau )A_{2}\left( \tau \right) L\right], 
\label{eq6}
\end{eqnarray}
where $G^{2}(\tau
)-g^{2}(\tau )=1$. After passing through the time lens, the photon will not only be amplified but also obtain the quadratic phase shift in time.

Similarly, after passing through the output dispersion medium of length $l$ \cite{PhysRevA.98.053815}
\begin{equation}
\hat{a}_{i}^{\left( 3\right) }\left( z+l, \tau \right) =\int_{-\infty }^{\infty
}G_{out}\left( \tau -\tau ^{\prime }\right) \hat{a}_{i}^{\left( 2\right)
}\left( z, \tau ^{\prime }\right) d\tau ^{\prime },
\label{shuchusesan}
\end{equation}
where $G_{out}\left( \tau \right) =\frac{1}{\sqrt{-2\pi iD_{out}}}e^{-i\tau
^{2}/2D_{out}}$, $D_{out}$ is the GDD of the output dispersive medium.

Under the condition of $D_{in}=-D_{out}=-D_{f}$, combined with Eqs.~(\ref{shurusesan}), (\ref{eq2}), (\ref{eq3}), (\ref{shuchusesan}), we get the input-output transformation relationship of the TF system based on the four-wave mixing time lens:
\begin{eqnarray}
\hat{a}_{s}\left( \tau \right)  &=&G(\tau )\hat{a}_{s}^{ in }\left( \tau
\right) +\frac{g(\tau )}{\sqrt{-iD_f}}\mathcal{F} [\hat{a}_i^{in\dagger}]\left( 
 \Omega=\frac{\tau}{D_f} \right),~~~~
\\
\hat{a}_{i}\left( \tau \right)  &=&G(\tau )\hat{a}_{i}^{ in }\left( \tau
\right)+\frac{g(\tau )}{\sqrt{- i D_f}}\mathcal{F}[\hat{a}_{s}^{ in \dagger }]\left(\Omega=\frac{\tau }{D_{f}}\right),~~~~
\end{eqnarray}
where \\
$\mathcal{F}[\hat{a}_{j}^{ in \dagger }]\left(\Omega=\frac{\tau }{D_{f}}\right)=\frac{1 }{\sqrt{2\pi}}\int \hat{a}_{j}^{in\dagger }\left( \tau ^{\prime }\right) \exp %
\left(-i \frac{\tau }{D_{f}} \tau ^{\prime }\right)
 d\tau ^{\prime }$, $j =\left\{ s,i\right\} $. For convenience, in the following part of this article $\tilde{a}_{j}^{ in \dagger }\left(\frac{\tau }{D_{f}}\right)$ is used to represent $\mathcal{F}[\hat{a}_{j}^{ in \dagger }]\left(\Omega=\frac{\tau }{D_{f}}\right)$. From the derived transformation relationship, it becomes evident that each time lens functions as a parametric amplifier with one arm amplifying the input and the other generating the Fourier transform of the input.

When there is only one strongly coherent input on the signal arm and the input idler field is in a vacuum state, the idler vacuum field can be ignored. At this time, the temporal intensity distribution of the output idler field is proportional to the Fourier transform of the input signal. Since the signal is not affected by the lens and is only affected by the positive and negative dispersion that cancel each other, resulting in perfect imaging of the signal, the output signal field is proportional to the input, as shown in the 
Ref.~\cite{2023PRLtemporalSU11}. However, when the signal input field is a weak field, such as a squeezed vacuum field, the vacuum input of the idler port becomes non-negligible.

\subsection{Temporal SU(1,1) interferometer}

We consider the temporal SU(1,1) interferometer shown in Fig.~\ref{fig1}, which consists of two TF systems. After the signal passes through TF$_1$, in the Fourier plane, we add a phase that changes linearly with time to the signal and idler, denoted as $\varphi _{1}=\beta _{1}+\gamma _{1}\tau $ and $\varphi _{2}=\beta _{2}+\gamma _{2}\tau $. And transmit them to TF$_2$, and the output of the interferometer is
\begin{eqnarray}
\hat{a}_{s}^{out}\left( \tau \right)  &=&g_{1}g_{2}\exp \left( -i\beta
_{2}\right) \sqrt{\frac{D_{f1}}{D_{f2}}}\hat{a}_{s}^{in}\left( \frac{D_{f1}}{%
D_{f2}}\tau +\gamma _{2}D_{f1}\right)   \nonumber \\
&+&G_{1}G_{2}\exp \left( i\varphi
_{1}\right)\hat{a}_{s}^{in}\left( \tau \right)  +g_{1}G_{2}\frac{\exp \left( i\varphi _{1}\right) }{\sqrt{%
-iD_{f1}}}  \nonumber \\
&\times &\tilde{a}_{i}^{in\dagger }\left( \frac{\tau }{D_{f1}}\right)
+G_{1}g_{2}\frac{\exp \left( -i\beta _{2}\right) }{\sqrt{-iD_{f2}}} 
\nonumber \\
&\times &\tilde{a}_{i}^{in\dagger }\left( \frac{\tau }{D_{f2}}+\gamma
_{2}\right) ,  \label{as_out}
\end{eqnarray}
\begin{eqnarray}
\hat{a}_{i}^{out}\left( \tau \right)  &=&g_{1}g_{2}\exp \left( -i\beta
_{1}\right) \sqrt{\frac{D_{f1}}{D_{f2}}}\hat{a}_{i}^{in}\left( \frac{D_{f1}}{%
D_{f2}}\tau +\gamma _{1}D_{f1}\right)   \nonumber \\
&+&G_{1}G_{2}\exp \left( i\varphi
_{2}\right)\hat{a}_{i}^{in}\left( \tau \right)  +g_{1}G_{2}\frac{\exp \left( i\varphi _{2}\right) }{\sqrt{%
-iD_{f1}}}  \nonumber \\
&\times &\tilde{a}_{s}^{in\dagger }\left( \frac{\tau }{D_{f1}}\right)
+G_{1}g_{2}\frac{\exp \left( -i\beta _{1}\right) }{\sqrt{-iD_{f2}}} 
\nonumber \\
&\times &\tilde{a}_{s}^{in\dagger }\left( \frac{\tau }{D_{f2}}+\gamma
_{1}\right) ,  \label{ai_out}
\end{eqnarray}
where $G_{i}$ and $g_{i}$ are the gain coefficients of the ${i}$th temporal lens, respectively, $D_{fi}$ is the focal GDD of the ${i}$th temporal lens, and $i=\{1,2\}$. In Eqs.~(\ref{as_out}) and ~(\ref{ai_out}), $\hat{a}_{s}^{in}\left( \tau \right)$ is the light of the input signal, and $\hat{a}_{i}^{in}\left( \tau \right)$ is the vacuum noise entering from the idle port, which is not considered in the temporal SU(1,1) theory in Ref.~\cite{2023PRLtemporalSU11}. The impact of noise on the system will be discussed in the next section.

Compared with the spatial SU (1,1) interferometer Ref.~\cite{2022Liang_Xinyun_Symmetry}, it is not difficult to find that the output of the temporal SU(1,1) interferometer is not just the coherent superposition of the input waveform after simple amplification. Instead, because the TF system performs a time Fourier transform, the output is the coherent superposition of the input waveform in one arm and the Fourier transformed waveform in the other arm, and both fields are amplified.

\section{Output spectrum with broadband squeezed light injection}

In this section, the broadband squeezed light is used as the input of the signal port of the temporal SU (1,1) interferometer and the characteristics of the output spectrum are analyzed. 

This light can be generated by an optical parametric amplifier (OPA) in a second-order nonlinear crystal \cite{1999RMP_Kolobov}, and its field operator is described by Fourier amplitude:
\begin{equation}
\hat{a}_{s}\left( \Omega \right) =\int \hat{a}_{s}\left( \tau \right)
e^{i\Omega \tau }d\tau,
\end{equation}
where $\Omega=\omega-\omega_0$ is the angular frequency shift and $\omega_0$ is the center frequency.
The generation of broadband squeezed light is given by the Bogoliubov transformation of the vacuum operator at the OPA input $(z=0)$ to the output $(z=l)$:
\begin{eqnarray}
\hat{a}_{s}\left( \Omega ,l\right) =U\left( \Omega \right) \hat{a}_{s}\left(
\Omega ,0\right) +V\left( \Omega \right) \hat{a}_{s}^{\dagger }\left(
-\Omega ,0\right), \label{bogoliubov}
\end{eqnarray}
where $U\left( \Omega \right) $ and $V\left( \Omega \right) $ are complex coefficients that depend on the OPA parameter gain and its phase matching condition in the nonlinear crystal \cite{1999RMP_Kolobov}. The transformation in Eq.~(\ref{bogoliubov}) describes the OPA's
production of a broadband squeezed vacuum at its output from a broadband
vacuum at its input.

Injecting $\hat{a}_{s}\left( \Omega ,l\right)$ into the temporal SU(1,1) interferometer signal port and vacuum state into another port, the two output fields are described:
\begin{eqnarray}
&&\hat{a}_{s}^{out}\left( \Omega \right) =G_{1}G_{2}\exp \left( i\beta
_{1}\right) \tilde{a}_{s}^{in}\left( \Omega -\gamma _{1}\right) +g_{1}g_{2}%
\sqrt{M}  \notag \\
&&\times \exp \left( -i\beta _{2}\right) \exp \left( i\gamma
_{2}D_{f2}\Omega \right) \tilde{a}_{s}^{in}\left( M\Omega \right) \notag \\
&&+G_{1}g_{2}\sqrt{ iD_{f2}}\exp \left( -i\beta _{2}\right) \exp \left(
i\gamma _{2}D_{f2}\Omega \right) \hat{a}_{i}^{in\dagger }\left(
-D_{f2}\Omega \right)  \notag\\
&&+g_{1}G_{2}\sqrt{ iD_{f1}}\exp \left( i\beta _{1}\right) \hat{a}%
_{i}^{in\dagger }\left( -D_{f1}\Omega +D_{f1}\gamma _{1}\right) ,  \label{as_out_omega}
\end{eqnarray}
and%
\begin{eqnarray}
&&\hat{a}_{i}^{out}\left( \Omega \right) =G_{1}G_{2}\exp \left( i\beta
_{2}\right) \tilde{a}_{i}^{in}\left( \Omega -\gamma _{2}\right) +g_{1}g_{2}%
\sqrt{M}  \notag \\
&&\times \exp \left( -i\beta _{1}\right) \exp \left( i\gamma
_{1}D_{f2}\Omega \right) \tilde{a}_{i}^{in}\left( M\Omega \right)  \notag\\
&&+G_{1}g_{2}\sqrt{ iD_{f2}}\exp \left( -i\beta _{1}\right) \exp \left(
i\gamma _{1}D_{f2}\Omega \right) \hat{a}_{s}^{in\dagger }(-D_{f2}\Omega ) \notag\\
&&+g_{1}G_{2}\sqrt{ iD_{f1}}\exp \left( i\beta _{2}\right) \hat{a}%
_{s}^{in\dagger }\left( -D_{f1}\Omega +\gamma _{2}D_{f1}\right) ,  \label{ai_outomega}
\end{eqnarray}
where $M=D_{f2}/D_{f1}$ is defined as scaling factor. 

Using a strongly monochromatic local oscillator to perform a balanced homodyne detection, we can measure the output fields. 
The measured quadrature operator is
\begin{equation}
\hat{X}_{j} =\hat{a}_{j}^{out} e^{-i\varphi }+%
\hat{a}_{j}^{out\dagger } e^{i\varphi }, ~j =\left\{ s,i\right\},
\end{equation}%
where $\varphi$ is the phase of the local light. Since the output field is the coherent superposition of the input waveform in one arm and the other arm after the Fourier transformation, and the input in the idler field is vacuum, the detection of the homodyne of the signal field and the idler frequency field use the frequency domain of the signal output $\hat{a}_{s}^{out}(\Omega)$ and the time domain of the idle output $\hat{a}_{i}^{out}(\tau)$, respectively.

\subsection{Signal output spectrum}
The measured quadrature operator of signal field is
\begin{equation}
\hat{X}_{s}\left( \Omega \right) =\hat{a}_{s}^{out}\left( \Omega \right) e^{-i\varphi }+%
\hat{a}_{s}^{out\dagger }\left( -\Omega \right) e^{i\varphi },
\end{equation}%
where $\varphi $ is the phase of the local light and the commutation
relation $[\hat{a}\left( \Omega \right) ,\hat{a}^{\dagger }\left( \Omega
^{\prime }\right) ]=2\pi \delta (\Omega -\Omega ^{\prime })$. The squeezed
spectrum $S\left( \Omega \right) $ is defined as $\left\langle \hat{X}%
^{\dagger }\left( \Omega \right) \hat{X}\left( \Omega ^{\prime }\right)
\right\rangle =2\pi S\left( \Omega \right) \delta \left( \Omega +\Omega
^{\prime }\right) $. 

For convenience, let $\beta _{1}$, $\beta _{2}$ and $\gamma _{2}$ be 0, that is, only the phase is added to the signal arm. Then, the output spectrum function of the interferometer signal port is
\begin{eqnarray}
S_{s}\left( \Omega \right)  &=&G_{1}^{2}G_{2}^{2}S_{in}\left( \Omega -\gamma
_{1}\right) +g_{1}^{2}g_{2}^{2}\left \vert M\right \vert S_{in}\left( M\Omega \right)   \notag \\
&&+g_{1}^{2}G_{2}^{2}+G_{1}^{2}g_{2}^{2}.
\end{eqnarray}%
$S_{s}\left( \Omega \right)$ consists of three terms. The first term is the center frequency shift term $G_{1}^{2}G_{2}^{2}S_{in}\left( \Omega -\gamma_{1}\right)$, where the phase derivative induces a shift in the spectrum. The frequency after the shift becomes $\omega'=\omega_0+\gamma_1$. The second term is the bandwidth scaling term $g_{1}^{2}g_{2}^{2}|M|S_{in}\left( M\Omega \right)$, which is unique to the temporal SU(1,1) interferometer. For any scaling factor $M$, there exists a relationship between the squeezing spectrum bandwidth $\Omega$ and the squeezing spectrum bandwidth $\Omega_{in}$ in the OPA output, given by $\Omega=\Omega_{in}/|M| $. The third term is the noise constant term $g_{1}^{2}G_{2}^{2}+G_{1}^{2}g_{2}^{2}$, where these terms are independent of the input light and reflect the noise contribution inherent in the interferometer system.

\begin{figure}[ht!]
\begin{center}
\centering{\includegraphics[scale=0.3,angle=0]{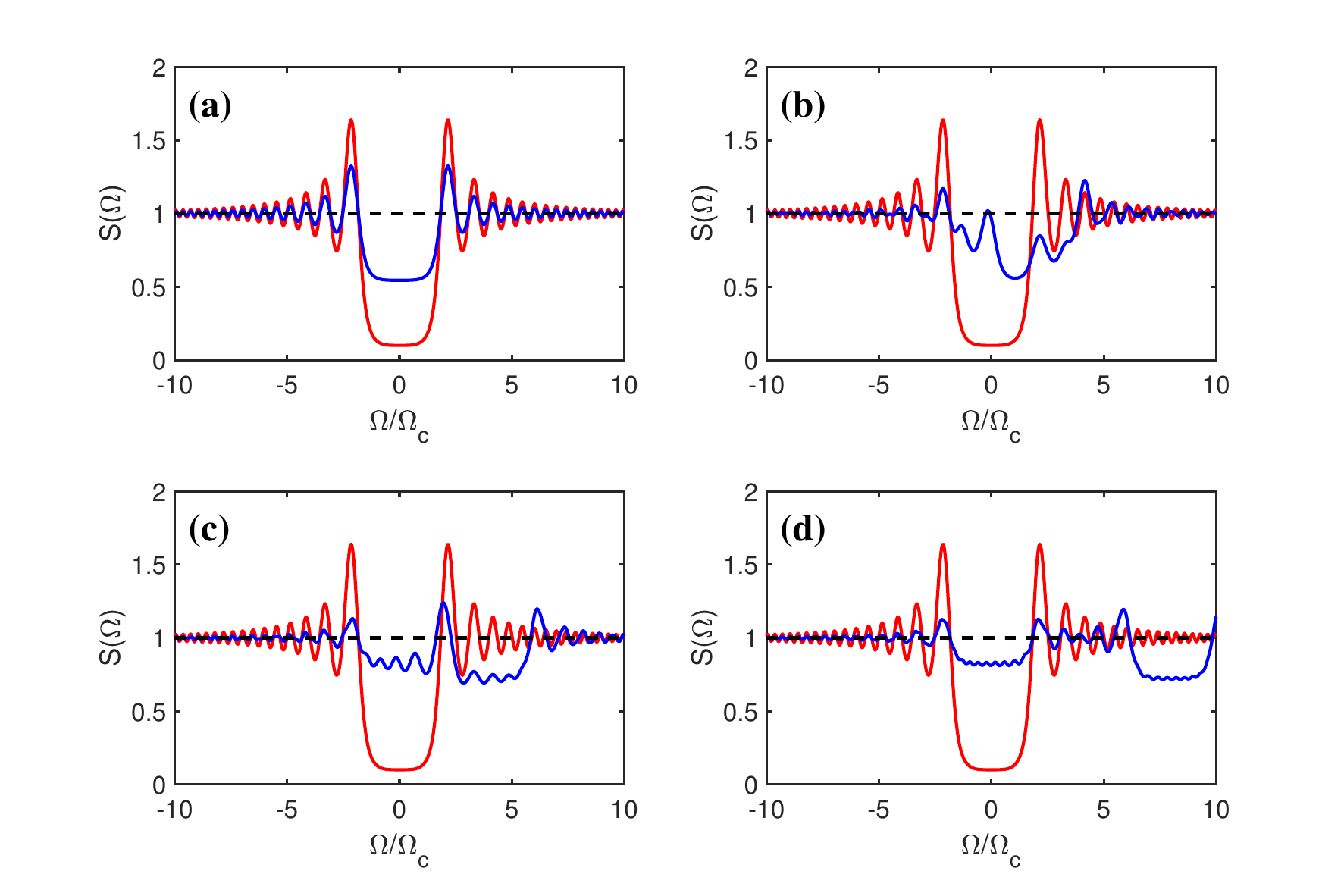}}
\end{center}
\caption{Squeezing spectra of a temporal SU(1,1) interferometer under the condition of broadband squeezed light injection with different phase derivatives. The blue line (red line) is the output (input) squeezing spectrum. The horizontal axis is scaled in units of $\Omega_c$ ($\Omega_c=1/\sqrt{\beta^{(2)}l}$), which represents the characteristic frequency of the broadband squeezed light output from the optical parametric amplifier. All spectral functions are normalized to the shot-noise level (black dotted lines). }\label{fig2}
\end{figure}

Here, we consider that the gain coefficients of the two time lenses are equal, that is, $G_{1}=G_{2}=G$ and $g_{1}=g_{2}=g$. When the scaling factor $M$ remains unchanged, it can be seen from Eq.~(17) that the larger the value of $G$, the more balanced the distribution of squeezing between the center frequency term and the frequency shift term. Next, we analyze the effect of the scaling factor $M$ on the output spectrum. When $M=1$, we discuss the influence of the phase derivative $\gamma_1$ of the phase on the spectrum function. When the frequency shift $\gamma_1$ of the center frequency is less than the squeezing bandwidth $\Omega_q$, the symmetry about the center frequency $\omega_0$ is destroyed, and the squeezing bandwidth of the output squeezing spectrum begins to increase. However, the squeezing performance is sacrificed due to the entry of vacuum noise. As shown in Fig.~\ref{fig2}b, the center frequency of $S_{out}$ becomes $\omega_0+\gamma_1/2$. As long as the center frequency of the output photocurrent is measured, the value of the phase derivative $\gamma_1$ can be obtained. 

\begin{figure}[ht!]
\begin{center}
\centering{\includegraphics[scale=0.48,angle=0]{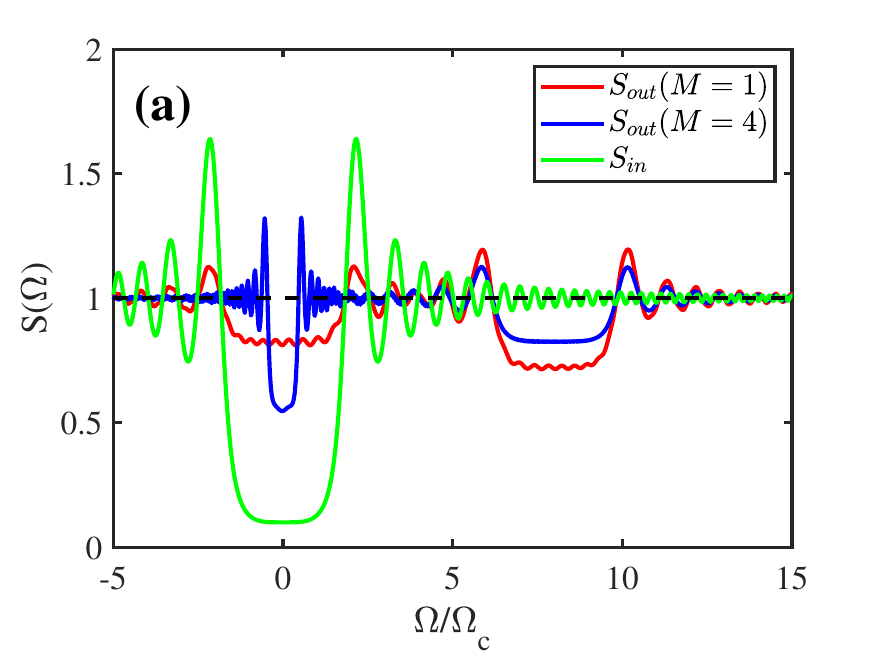}}
\centering{\includegraphics[scale=0.48,angle=0]{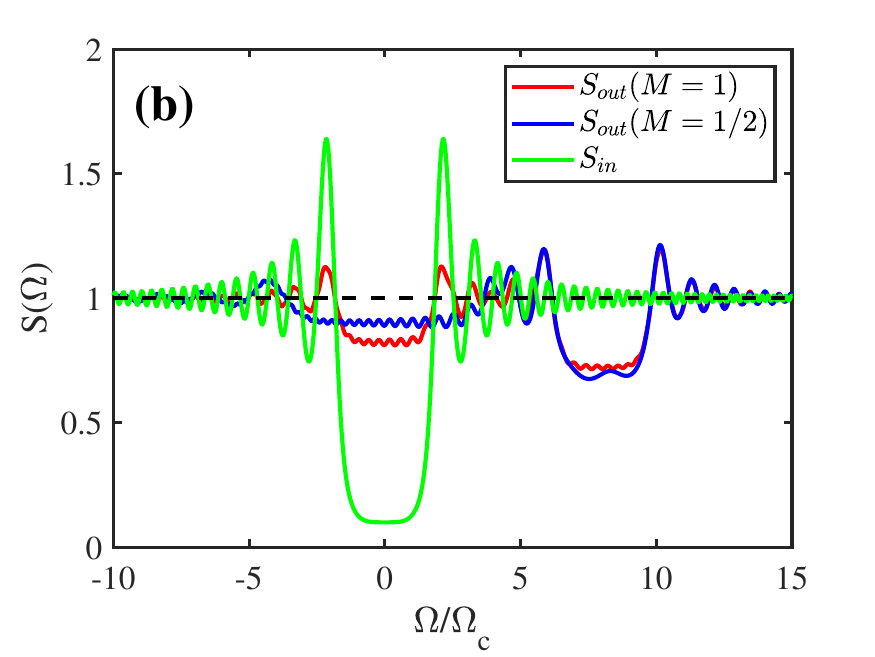}}
\end{center}
\caption{Squeezing spectra of a temporal SU(1,1) interferometer under the condition of broadband squeezed light injection with different scaling factors. The horizontal axis is scaled in units of $\Omega_c$ ($\Omega_c=1/\sqrt{\beta^{(2)}l}$). All spectral functions are normalized to the shot-noise level (black dotted lines).}\label{fig3}
\end{figure}

As $\gamma_1$ increases to be equal to the squeezing bandwidth $\Omega_q$, the squeezing bandwidth stops increasing, and two frequency bands begin to appear where the quantum fluctuations of the photocurrent are lower than the shot noise level, and the squeezing bandwidth of each frequency band is equal to the squeezing bandwidth of the input light, as shown in Fig.~\ref{fig2}c. As $\gamma_1$ continues to increase, the shift effect will become more obvious. From a physical point of view, the adjustment of the phase derivative $\gamma_1$ provides a flexible means to control the center frequency position of the squeezing spectrum, thereby achieving selective regulation of signals in a specific frequency band. However, excessive shift may lead to a decrease in the effective utilization of the squeezing bandwidth. Therefore, in practical applications, $\gamma_1$ should be reasonably selected according to the signal frequency distribution and target requirements to achieve optimal coverage of the squeezing spectrum. The results of this paper show that spectrum shift provides an important reference for exploring the combination of broadband squeezing light and frequency selective applications.
\par
When $M\neq1$, the term $g_{1}^{2}g_{2}^{2}|M|S_{in}\left( M\Omega \right)$ will have an impact on the output spectrum. By changing the value of $M$, the dominant role of the frequency shift term and the scaling term in $S_{t}$ can be adjusted. As $M$ increases, the degree of squeezing at the central frequency increases while the bandwidth narrows. Meanwhile, since the scaling term becomes predominant, the squeezing performance at $\omega_0+\gamma_1$ reduces. In the high-amplification limit where $M\gg1$, the influence of the shift term can be ignored.

When $M<1$, the spectrum shift term is dominant, and the squeezing behavior of the output spectrum is just the opposite. 
As shown in Fig.~\ref{fig3}(b), the degree of squeezing of the output spectrum decreases at position $\Omega=0$, while at $\omega_0+\gamma_1$ the degree of squeezing allocated constitutes a majority proportion. The squeezing spectrum of the interferometer is concentrated in a spectral interval, making it easier to match it with a specific detector or system in practical applications. For example, optical detectors are usually optimized for a certain frequency band, and this single interval characteristic is easier to meet the needs. When $M<0$ and the input spectrum is symmetric, given that the scaling factor of output spectrum is related to $|M|$, the behavior of the output spectrum for $M<0$ is similar to that for $M>0$.

\subsection{Idler output spectrum}

From the Eq.~(\ref{ai_out}), we can see that the first two terms of the output idler field are proportional to the input idler field, and the last two terms are proportional to the Fourier transform of the input signal field, which is still a function of time. There is a mapping between the frequency of the signal arm and the time of the idler arm, \(\Omega = \tau / D_{f1}\) and \(\Omega = \tau / D_{f2}+\gamma_1\). That is, the frequency domain information of the input signal light is mapped to the time domain waveform. Then, the quantum noise characteristics in the frequency domain can be observed directly through temporal measurement. Thus, the quadrature operator of idler field is
\begin{equation}
  \hat{X}_{i}(\tau)=\hat{a}_{i}^{out}(\tau)e^{-i\varphi}+\hat{a}_{i}^{out\dagger}(\tau)e^{i\varphi}.
\end{equation}

Using the same conditions as those for the signal port, by setting \(\beta_1\), \(\gamma_2\) and \(\beta_2\) to 0, the output squeezed spectrum function at the idler port is obtained as
\begin{eqnarray}
S_{i}\left( \tau \right)  &=&\frac{1}{\left\vert D_{f1}\right\vert }%
g_{1}^{2}G_{2}^{2}S_{in}\left( \frac{\tau }{D_{f1}}\right)+\frac{1}{\left\vert MD_{f1}\right\vert }G_{1}^{2}g_{2}^{2}\nonumber\\
&\times&S_{in}\left(
\frac{\tau }{M D_{f1}}+\gamma _{1}\right)+G_{1}^{2}G_{2}^{2}+\frac{g_{1}^{2}g_{2}^{2}}{\left\vert M\right\vert }. 
\end{eqnarray}

Similarly to the composition of the signal output spectrum, the idler output spectrum is also composed of three terms.
The first term is \(g_{1}^{2}G_{2}^{2}S_{in}\left(\tau/D_{f1}\right)/{|D_{f1}|}\), which reveals the important conversion characteristics of the frequency-to-time domain of the temporal SU (1,1) interferometer. Specifically, the interferometer can accurately map the input spectrum \(S_{in}(\Omega)\) from the frequency domain \(\Omega=\tau / D_{f1}\) to the time domain. This process is of key significance in the field of optical signal processing and spectrum analysis, and provides theoretical support for a deep understanding of the conversion mechanism of signals between the time and frequency domains. The second term
\(G_{1}^{2}g_{2}^{2}S_{in}\left({\tau}/({MD_{f1}})+\gamma_1\right)/{|MD_{f1}|}\) belongs to the phase-shift term. Similarly to the first term, it also undergoes mapping and scaling operations. It should be noted that the introduction of the phase derivative \(\gamma_1\) has a unique effect, which can shift the input spectrum in the time domain. From the perspective of the frequency domain, this translation operation is equivalent to moving the center frequency \(\Delta\omega = \gamma_1\). This feature has an important application value in spectrum control and signal processing and can be used to achieve precise positioning and adjustment of the spectrum. The third term is the noise constant term, which is expressed as \(G_{1}^{2}G_{2}^{2}+{g_{1}^{2}g_{2}^{2}}/{|M|}\). This constant term is determined by the gain coefficients $G_{1}$ and $G_{2}$ of the two time lenses and the amplification coefficient \(|M|\). In the optical system and signal processing process, the noise constant term is one of the key factors affecting the system performance and signal quality. 

\begin{figure}[ht!]
\begin{center}
\centering{\includegraphics[scale=0.48,angle=0]{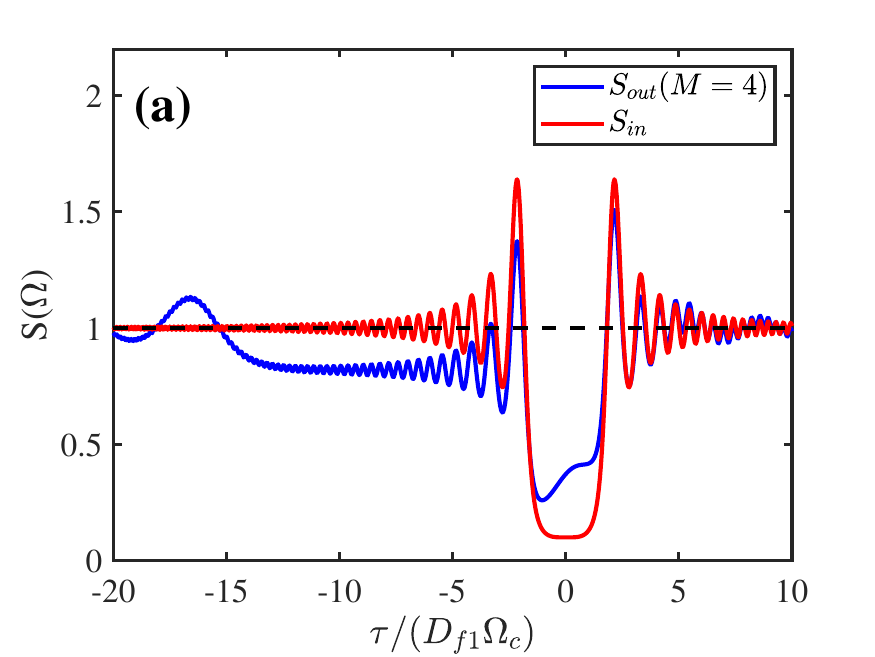}}
\centering{\includegraphics[scale=0.48,angle=0]{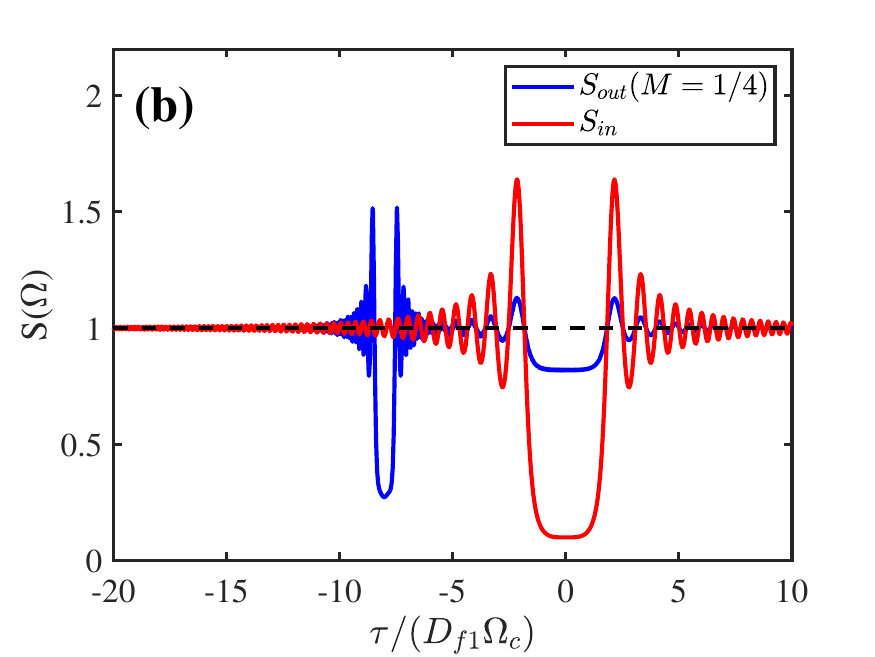}}
\end{center}
\caption{Squeezing spectra at the idler output port of the temporal SU(1,1) interferometer under broadband squeezed light injection for different scaling factors. The horizontal axis is mapped from the time domain to the frequency domain and is scaled in units of \(\Omega_c\) ($\Omega_c=1/\sqrt{\beta^{(2)}l}$). All spectral functions are normalized to the shot-noise level (black dotted lines).}\label{fig4}
\end{figure}

Next, assuming that $G_{1}=G_{2}$, $\gamma_1 = 8\omega_{c}$, when $M= 1$, due to the control of \(\gamma_1\), the squeezed spectrum output similar to Fig.~\ref{fig2}(d) is divided into two parts. Unlike the signal field, the shift caused by the phase derivative moves toward the low-frequency direction. When \(M > 1\), as shown in Fig.~\ref{fig4}(a), the bandwidth of the translation term will be widened \(M\) times. At the same time, the squeezed performance will also decrease accordingly. However, the term vacuum noise will be suppressed to \(1/|M|\) of its original value, thus reducing the interference of vacuum noise on the squeezed section at the center frequency. This shows that when \(M > 1\), the SU(1,1) interferometer can achieve broadband quantum squeezing by adjusting the amplification factor. When \(M < 1\), as shown in Fig.~\ref{fig4}(b), the bandwidth of the translation term is squeezed, but its squeezed degree is correspondingly improved. However, the noise term \({g_{1}^{2}g_{2}^{2}}/{|M|}\) is correspondingly amplified, resulting in a decrease in the compression performance of the term.

\section{Conclusion}
In conclusion, we study the temporal SU(1,1) interferometer based on a temporal Fourier transform system, and consider the influence of vacuum noise in the temporal SU(1,1) interferometer. In the case of non-classical light injection such as broadband squeezed light, by performing homodyne measurement on the output fields, we found that the interferometer is sensitive to the phase derivative and scaling factor. The phase derivative induces a frequency shift effect, where the magnitude of this shift exhibits a linear relationship with the phase derivative. In the output squeezed-state spectrum, the distribution of squeezing degree concentrates at the center frequency and locations where frequency shifts occur. As scaling factor $M$ increases, the proportion of squeezing degree allocated at the center frequency correspondingly increases. This result can adjust the spectral distribution of the input signal, and then selectively detect the squeezing characteristics of a specific frequency band. It can also be used to compensate for the spectral mismatch of system components so that the signal matches the corresponding bandwidth of the detector.

\section{ACKNOWLEDGMENTS}

This work is supported by the Shanghai Science and Technology Innovation Project No. 24LZ1400600, the Innovation Program for Quantum Science and Technology 2021ZD0303200; the National Natural Science Foundation of China Grants No. 11974111 and Fundamental Research Funds for the Central Universities.

\appendix
\section{Output spectrum functions}
Using a strongly monochromatic local oscillator to perform a balanced homodyne detection, we can measure the output fields. The measured quadrature operator is
\begin{equation}
\hat{X}_{j}\left( \Omega \right) =\hat{a}_{j}\left( \Omega \right) e^{-i\varphi }+%
\hat{a}_{j}^{\dagger }\left( -\Omega \right) e^{i\varphi },~j=\{s,i\}
\end{equation}%
where $\varphi $ is the phase of the local light and the commutation
relation $[\hat{a}\left( \Omega \right) ,\hat{a}^{\dagger }\left( \Omega
^{\prime }\right) ]=2\pi \delta (\Omega -\Omega ^{\prime })$. The squeezed
spectrum $S\left( \Omega \right) $ is defined as $\left\langle \hat{X}%
^{\dagger }\left( \Omega \right) \hat{X}\left( \Omega ^{\prime }\right)
\right\rangle =2\pi S\left( \Omega \right) \delta \left( \Omega +\Omega
^{\prime }\right) $. 
In this appendix, we give the general spectrum functions of output signal and idler fields:
\begin{widetext}
\begin{eqnarray}
S_{s}\left( \Omega \right)  &=&G_{2}^{2}G_{1}^{2}\left\langle \tilde{a}%
_{s}^{in\dagger }\left( -\Omega +\gamma _{1}\right) \tilde{a}_{s}^{in}\left(
\Omega -\gamma _{1}\right) \right\rangle +G_{2}^{2}G_{1}^{2}\exp \left(
2i\varphi -2i\beta _{1}\right) \left\langle \tilde{a}_{s}^{in\dagger }\left(
-\Omega +\gamma _{1}\right) \tilde{a}_{s}^{in\dagger }\left( -\Omega +\gamma
_{1}\right) \right\rangle  \nonumber\\
&&+G_{2}^{2}G_{1}^{2}\exp \left( -2i\varphi +2i\beta _{1}\right)
\left\langle \tilde{a}_{s}^{in}\left( \Omega -\gamma _{1}\right) \tilde{a}%
_{s}^{in}\left( \Omega -\gamma _{1}\right) \right\rangle
+G_{2}^{2}G_{1}^{2}\left\langle \tilde{a}_{s}^{in}\left( \Omega -\gamma
_{1}\right) \tilde{a}_{s}^{in\dagger }\left( -\Omega +\gamma _{1}\right)
\right\rangle  \nonumber\\
&&+g_{1}^{2}g_{2}^{2}\left\vert M\right\vert \left\langle \tilde{a}%
_{s}^{in\dagger }\left( -M\Omega \right) \tilde{a}_{s}^{in}\left( M\Omega
\right) \right\rangle  +g_{1}^{2}g_{2}^{2}\left\vert M\right\vert \left\langle \tilde{a}%
_{s}^{in}\left( M\Omega \right) \tilde{a}_{s}^{in\dagger }\left( -M\Omega
\right) \right\rangle \nonumber\\
&&+g_{1}^{2}g_{2}^{2}\left\vert M\right\vert \exp \left( 2i\beta
_{2}-2i\gamma _{2}D_{f2}\Omega +2i\varphi \right) \left\langle \tilde{a}%
_{s}^{in\dagger }\left( -M\Omega \right) \tilde{a}_{s}^{in\dagger }\left(
-M\Omega \right) \right\rangle  \nonumber\\
&&+g_{1}^{2}g_{2}^{2}\left\vert M\right\vert \exp \left( -2i\beta
_{2}+2i\gamma _{2}D_{f2}\Omega -2i\varphi \right) \left\langle \tilde{a}%
_{s}^{in}\left( M\Omega \right) \tilde{a}_{s}^{in}\left( M\Omega \right)
\right\rangle+g_{1}^{2}G_{2}^{2}+g_{2}^{2}G_{1}^{2}, \label{Sout_s}
\end{eqnarray}
\end{widetext}
and
\begin{widetext}
\begin{eqnarray}
S_{i}\left( \tau \right)  &=&g_{1}^{2}G_{2}^{2}\exp \left( 2i\varphi
_{2}-2i\varphi +i\pi /2\right) \frac{1}{\left\vert D_{f1}\right\vert }%
\left\langle \tilde{a}_{s}^{in\dagger }\left( \frac{\tau }{D_{f1}}\right) 
\tilde{a}_{s}^{in\dagger }\left( \frac{\tau }{D_{f1}}\right) \right\rangle \nonumber
\\
&&+g_{1}^{2}G_{2}^{2}\exp \left( -2i\varphi _{2}+2i\varphi -i\pi /2\right) 
\frac{1}{\left\vert D_{f1}\right\vert }\left\langle \tilde{a}_{s}^{in}\left(
-\frac{\tau }{D_{f1}}\right) \tilde{a}_{s}^{in}\left( -\frac{\tau }{D_{f1}}%
\right) \right\rangle  \nonumber\\
&&+g_{1}^{2}G_{2}^{2}\frac{1}{\left\vert D_{f1}\right\vert }\left\langle 
\tilde{a}_{s}^{in}\left( -\frac{\tau }{D_{f1}}\right) \tilde{a}%
_{s}^{in\dagger }\left( \frac{\tau }{D_{f1}}\right) \right\rangle
+g_{1}^{2}G_{2}^{2}\frac{1}{\left\vert D_{f1}\right\vert }\left\langle 
\tilde{a}_{s}^{in\dagger }\left( \frac{\tau }{D_{f1}}\right) \tilde{a}%
_{s}^{in}\left( -\frac{\tau }{D_{f1}}\right) \right\rangle  \nonumber\\
&&+G_{1}^{2}g_{2}^{2}\exp \left( -2i\beta _{1}-2i\varphi +i\pi /2\right) 
\frac{1}{\left\vert D_{f2}\right\vert }\left\langle \tilde{a}_{s}^{in\dagger
}\left( \frac{\tau }{D_{f2}}+\gamma _{1}\right) \tilde{a}_{s}^{in\dagger
}\left( \frac{\tau }{D_{f2}}+\gamma _{1}\right) \right\rangle  \nonumber\\
&&+G_{1}^{2}g_{2}^{2}\exp \left( 2i\beta _{1}+2i\varphi -i\pi /2\right) 
\frac{1}{\left\vert D_{f2}\right\vert }\left\langle \tilde{a}_{s}^{in}\left(
-\frac{\tau }{D_{f2}}-\gamma _{1}\right) \tilde{a}_{s}^{in}\left( -\frac{%
\tau }{D_{f2}}-\gamma _{1}\right) \right\rangle  \nonumber\\
&&+G_{1}^{2}g_{2}^{2}\frac{1}{\left\vert D_{f2}\right\vert }\left\langle 
\tilde{a}_{s}^{in}\left( -\frac{\tau }{D_{f2}}-\gamma _{1}\right) \tilde{a}%
_{s}^{in\dagger }\left( \frac{\tau }{D_{f2}}+\gamma _{1}\right)
\right\rangle  \nonumber\\
&&+G_{1}^{2}g_{2}^{2}\frac{1}{\left\vert D_{f2}\right\vert }\left\langle 
\tilde{a}_{s}^{in\dagger }\left( \frac{\tau }{D_{f2}}+\gamma _{1}\right) 
\tilde{a}_{s}^{in}\left( -\frac{\tau }{D_{f2}}-\gamma _{1}\right)
\right\rangle +G_{1}^{2}G_{2}^{2}+g_{1}^{2}g_{2}^{2}\left\vert \frac{D_{f1}}{%
D_{f2}}\right\vert \label{Sout_i}
\end{eqnarray}
\end{widetext}

Considering the phase is only added to the signal arm, i.e. $\beta_2=0$ and $\gamma_2= 0$, and for convenience $\beta_1=0$, then Eqs.~(\ref{Sout_s}) and (\ref{Sout_i}) can be simplified as
\begin{eqnarray}
S_{s}\left( \Omega \right)  &=&G_{1}^{2}G_{2}^{2}S_{in}\left( \Omega -\gamma
_{1}\right) +g_{1}^{2}g_{2}^{2}\left \vert M\right \vert S_{in}\left( M\Omega \right)   \notag \\
&&+g_{1}^{2}G_{2}^{2}+G_{1}^{2}g_{2}^{2},
\end{eqnarray}
and
\begin{eqnarray}
S_{i}\left( \tau \right)  &=&\frac{1}{\left\vert D_{f1}\right\vert }%
g_{1}^{2}G_{2}^{2}S_{in}\left( \frac{\tau }{D_{f1}}\right)+\frac{1}{\left\vert MD_{f1}\right\vert }G_{1}^{2}g_{2}^{2}\nonumber\\
&\times&S_{in}\left(
\frac{\tau }{M D_{f1}}+\gamma _{1}\right)+G_{1}^{2}G_{2}^{2}+\frac{g_{1}^{2}g_{2}^{2}}{\left\vert M\right\vert }. 
\end{eqnarray}

From the above equation of the output spectrum, when broadband squeezed light is injected into the input port of the signal light, the signal light spectrum at the output end will present a unique form: it is composed of the superposition of the original signal light and the signal light that has undergone frequency shift in the frequency domain. At the same time, the output spectrum of the idler light also shows similar characteristics, but its essence corresponds to the inverse Fourier transform of the input signal light in the time domain, and this spectral characteristic is still a function of frequency.

\bibliography{template-2024} %%%% bib file name
\nolinenumbers
\end{document}